\documentstyle[aps,prl,epsfig,floats]{revtex}
\def\sige{Si$_x$Ge$_{1-x}$}
\def\sic{Si$_x$C$_{1-x}$}
\def\sio2{$\mathrm{SiO_2}$}
\def\si29{$^{29}$Si}
\def\si{Si}

\def\mob{cm$^2$/Vsec}

\def\mub{\mu _B}

\def\p31{$^{31}$P}

\def\tfi{t_\phi}

\def\ud{\uparrow \downarrow} 
\def\du{\downarrow \uparrow}
\def\dd{\downarrow \downarrow}

\def\st{|\ud +\du \rangle}
\def\ss{|\ud -\du \rangle}

\begin{document}
\twocolumn[
\hsize\textwidth\columnwidth\hsize\csname@twocolumnfalse\endcsname
\draft

\title{Silicon-based Quantum Computation}

\author{B. E. Kane\thanks{e-mail: kane@lps.umd.edu}}

\address{Laboratory for Physical Sciences\\
University of Maryland\\
College Park, MD 20740 USA }

\date{\today}

\maketitle

\begin{abstract}
An architecture for a quantum computer is presented in which
spins associated with donors in silicon function as qubits. 
Quantum operations on the spins are performed using a
combination of voltages applied to gates adjacent to the spins
and radio frequency magnetic fields applied resonant with spin
transitions.  Initialization and measurement of electron spins
is made by electrostatic probing of a two electron system, whose
orbital configuration must depend on the spin states of the
electrons because of the Pauli Principle. Specific devices will be discussed
which perform all the necessary operations for quantum computing,
with an emphasis placed on a qualitative presentation of the
principles underlying their operation.

The likely impediments to achieving large-scale
quantum computation using this architecture will
be addressed: the computer
must operate at extremely low temperature,
must be fabricated from devices built with near atomic precision,
and will require extremely accurate gating operations in order to
perform quantum logic.  Refinements
to the computer architecture will be presented
which could remedy each of these
deficiencies.  I will conclude by discussing a possible
specific realization of the computer using Si/\sige~ heterostructures 
into which donors are deposited using a low energy 
focused ion beam. 

\end{abstract}

\medskip
\medskip

]

\section{Introduction}

One of the greatest challenges facing physics at the onset of the
twenty-first century is to determine whether or not construction
of a large-scale quantum computer is possible both in principle
and in terms of practical technical issues. While quantum algorithms
have been developed which prove convincingly that quantum computers
have capabilities conventional computers cannot duplicate,
future applications of these hypothetical machines remain mostly
a subject of speculation.  It seems likely, however, that if a
large-scale quantum computer can be built, then these machines
will have a significant impact in the coming century, when information
processing technology will only be more important than it is
today. 

Large-scale \it conventional~\rm computation only became possible when
thousands of transistors could be integrated onto a single solid state
chip, and it is widely believed that scalable quantum computation
will be achieved when solid state quantum logical devices are
similarly integrated.  It should be stated at the outset, however,
that since the likely impediments to large-scale quantum computation
will be altogether different from those that faced conventional
computers, scaling a solid state quantum computer may not prove to
be viable.  Ultimately the only test of viability of any quantum
computer architecture must come from the laborious development and refinement
of experimental candidate devices and architectures.

While by no means assured of success, quantum computer architectures
implemented in solids, and especially in semiconductors, have the
potential to take advantage of the enormous amount of ingenuity and
resources that have gone into the development of contemporary
microelectronics.  This development will no doubt continue until
electronic device sizes approach atomic dimensions, a realm in which
a wide variety of device concepts potentially capable of quantum
computation will be possible to implement.

Spins are perhaps the ideal qubits embodied in natural
systems, and several candidate quantum computer architectures have
been developed with spin qubits embedded in solid state materials
\cite{Loss98} \cite{Privman98} \cite{Kane98} 
\cite{Imamoglu99} \cite{Vrijen99} \cite{Bandyopadhyay99}.
The essence of this author's published proposal for a solid state
quantum computer \cite{Kane98} is that spins associated with donors in silicon
- the semiconductor that is the mainstay of conventional computer
technology - are ideally suited to function as qubits in a quantum
computer.  This fact is a consequence of the high degree of isolation
of electron and nuclear spins at donors in Si from their surroundings
at low temperatures ($T<$4K).  This isolation means that the highly
coherent manipulations of the spins necessary for quantum computation
are possible if there is a suitable means of controlling the interactions
of individual spins with each other and with externally applied fields.
In the proposal \cite{Kane98},
control of individual spin dynamics is achieved by applying
voltages to metal gates, located adjacent to the donors.  The donors must
be in prescribed locations within the silicon host to enable 
accurate gate control, a requirement which will pose perhaps the most
significant obstacle to the implementation of the computer design.

Readout of single spin qubits (and also initialization of the qubits)
is achieved by a two step process: first, a spin quantum number is
transferred to a property of the charge configuration of a system.
Second, the charge configuration is determined using sensitive
electronic devices.  Spin quantum numbers can most easily affect
the charge configuration of a two electron system: the Pauli
Exclusion Principle requires that two electrons can only occupy
the same orbital state if they are in a mutual spin singlet state,
with spins pointing in opposite directions.  Sensitive electrometers
now have resolution much better than a single electron charge, and
are thus capable of determining whether two electrons are in a
singlet or a triplet state (when the two spins are pointing
in the same direction and the electrons cannot occupy the same
orbital state).

The computer is capable of performing both logic 
and measurement operations in parallel and
is thus compatible with error correcting algorithms, which will
be an inevitable attribute of any large-scale quantum computer.
The high degree of isolation of spins in Si means that performing
logical operations with the accuracy required ($\leq 10^{-4}$)
for error corrected
continuous computation \cite{Knill98} \cite{Gottesman99}
is in principle possible.
  
The architecture is not without its deficiencies,
and subsequent to the publication of
the original proposal several suggestions were made that could
lead to improvements of the computer design \cite{Loss99},
\cite{Benjamin00},\cite{Vrijen99}.  
Consequently, I will address several specific modifications that will likely
need to be incorporated into the architecture if it is to be scaled:
firstly, I will discuss
techniques that would allow the computer to operate at higher temperatures
by refrigerating spins on the computer chip.  Secondly, I will describe
possible ways in which free electrons can be used
to transmit quantum information across large distances on the computer
chip \cite{Loss99}, making the execution of quantum algorithms more efficient.
Finally, I will discuss approaches to quantum logic that are insensitive to the
inevitable fluctuations present in solid state devices \cite{Benjamin00}.

I will conclude by presenting a specific material and fabrication
technology for the computer: high quality Si/\sige~ heterostructures
grown by molecular beam epitaxy into which single donor ions are
deposited during heterostructure growth \cite{Vrijen99}. The devices and fabrication
technologies presented, however, are not
intended to be a blueprint for the construction of a quantum computer but rather are
an effort to stimulate research and thinking both in the nanostructure
and device physics community, as well as in the community focused on
theoretical issues of quantum computation and measurement.

\section{Electron and Nuclear Spins at Donors in Silicon}

For crystals of pure Si at low temperatures ($T$), no electrons are in the
conduction band, and the material is an insulator.  The addition of
Column V donors to the Si crystal results in electron states near in
energy to the conduction band but weakly bound to the donor sites at
low $T$.  The theory of these weakly bound states was developed
in the fifties \cite{Kohn}; the electron states resemble those in a hydrogen
atom, but with an expanded Bohr radius ($a_B \cong $15-30 \AA) 
and a reduced binding energy ($E_b \cong$ 10-50 meV).

\begin{figure}
\vspace{0cm}
\begin{center}
\includegraphics[width=8.5cm,angle=0]{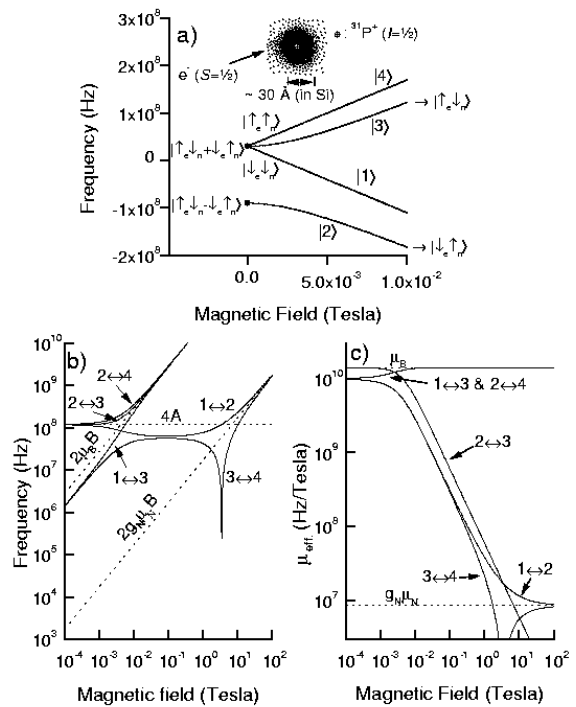}
\end{center}
\vspace{0cm}
\caption{(a) Spin energy levels of a \p31 donor in Si and
(b) energy differences of the levels as a function of applied
magnetic field, $B$.  At $B$=0 the electron-nuclear spin singlet
is 120 MHz below the degenerate triplet states.  As $B$ is
increased these energy differences ultimately approach the
electron and nuclear Zeeman energies.  (c) The matrix
elements coupling the different levels as a function of
$B$.  Coupling between states with different nuclear spin
decreases with increasing $B$, and consequently
operation of a nuclear spin quantum computer is
potentially
more rapid at smaller magnetic fields. }
\end{figure}

The electron has spin $S$=1/2, and all Column V donors have nonzero nuclear 
spin $I$.  (In contrast all Group IV elements,
including Si, have stable isotopes with
$I$=0, which means nuclear spins can in principle be entirely eliminated from
these materials by isotope refinement.)
The simplest system - and the one that has been most exhaustively
studied - is P doped Si, in which $I$=1/2.
The spin Hamiltonian for the nucleus-electron system in Si:P, with 
magnetic field $B\parallel z$ is:
\begin{equation}
\label{1}
H_{en} = \mub B \sigma_z^e -g_n \mu_n B \sigma_z^n + A 
                     {\bf{\sigma}}^e \cdot {\bf{\sigma}}^n  ,    
\end{equation}
where ${\bf{\sigma}}$ are the Pauli spin matrices (with eigenvalues $\pm$ 1), 
$\mu_B$ and $\mu_n$ are respectively the Bohr and nuclear
magneton, $g_n$  is the nuclear g-factor (=1.13 for \p31), and
$A= \frac{8}{3} \pi \mub g_n \mu_n  |  \Psi (0) |^2 $
is the contact hyperfine interaction energy, with $| \Psi(0) |^2$ the probability density of the
electron wave function evaluated at the nucleus \cite{Slichter}.  For electrons in Si,
$|g-2| \cong 10^{-3}$, and consequently $g$=2 is assumed in Eq. \ref{1}. 
Because the electron wave function is
strongly peaked at the donor site, the contact hyperfine interaction energy
of the electron and the donor nucleus
greatly exceeds dipolar spin interactions.  The Hamiltonian
(\ref{1}) can readily be solved exactly \cite{Feynman} and the energy levels for
Si:P are plotted in Fig. 1 as a function of $B$.

At $B$=0 energy eigenstates of Eq. \ref{1} are also eigenstates of the spin exchange
operator.  The ground state is the singlet 
($| \uparrow_e \downarrow_n -\downarrow_e \uparrow_n \rangle$), lying
4$A$ (=120 MHz in Si:P \cite{Feher59a}) below the threefold degenerate 
triplet excited states.
Application of $B$ leads to evolution of the states into well defined electron
and nuclear spin states as first the electron and then the nuclear Zeeman energies exceed $4A$
(Fig. 1b).  Transitions between the different energy levels are induced by
a radio frequency magnetic field $B_{rf}$ applied at a frequency resonant with the energy
level difference.  The magnitude of the matrix element $\mu_{eff.}$ coupling
between the levels is plotted in Fig.
1c.  At $B$=0, couplings for allowed transitions are approximately the same
as for an uncoupled electron spin, $\mu_B$.  As $B$ is increased and
electron and nuclear transitions become distinct, the coupling between the
nuclear states weakens, approaching $\mu_n$ at large $B$.

There are several reasons why the coupled electron-nucleus system at donors in Si
is an excellent building block for quantum computational devices: first,
the transition times between the states depicted in Fig. 1 can be exceedingly
long at low temperatures and are of order one hour at $T$=1 K and $B$=0.1 T
\cite{Feher59b}\cite{Honig60}. 
Secondly, the linewidths of the transitions are very narrow and are limited by
the small percentage of $^{29}$Si nuclei with $I$=1/2 present in the
Si crystal that also interact with the electron via hyperfine interactions \cite{Feher59b}.
In isotopically purified $^{28}$Si (with $I$=0), Si:P linewidths are $<$ 1 MHz
\cite{Feher59b}.  The
most relevant factor for qubits is the phase relaxation time, $t_\phi$.
In isotopically purified Si:P, $t_\phi$ exceeds 0.5 msec \cite{Gordon58} \cite{Chiba72},
and is probably limited by dipolar interactions between the electron spins.
In a quantum computer dipolar interactions can be eliminated as a source of decoherence using
compensating algorithms \cite{Viola99}.

\section{Quantum Operations with the \si:P System}

The Si:P system is a natural two qubit quantum computer
\cite{Divincenzo95b}:
the controlled NOT operation, in which an electron spin flip occurs
conditioned on the state of the nuclear spin, is performed by
exciting the transition 
$|\downarrow_e \uparrow_n \rangle \leftrightarrow | \uparrow_e \uparrow_n \rangle$, 
for example.  The SWAP operation is performed by exciting the
transition 
$|\downarrow_e \uparrow_n \rangle \leftrightarrow | \uparrow_e \downarrow_n \rangle$.
The electron and nuclear
spins have distinct characteristics that are favorable for quantum computation.
The electron spins are de-localized and mobile, and they can couple to
additional nuclei via the hyperfine interaction or to other electrons via
the exchange interaction.  The location of the electron can be controlled
by electric fields applied to gates on the Si surface, as is done
in conventional Si devices.  Electron spins can also in principle be
measured using device concepts based on the Pauli Principle. The nuclear
spins are much more weakly coupled to the environment, have no orbital
degrees of freedom, rotate much more slowly in an applied $B$
than electron spins, and are consequently almost ideally suited to
function as qubits in a quantum computer.

While a quantum computer architecture has been proposed based on donors in 
silicon-germanium heterostructures which uses exclusively the electron spins
\cite{Vrijen99}, no common
donor exists for these semiconductors that does not possess a nuclear spin.
Consequently, it is preferable to adopt an architecture which uses the
properties of both the electron and nuclear spins to advantage.  In the
architecture to be presented here, nuclear spins are the qubits and quantum
memory of the computer, while the electrons are used to $mediate$ interactions
between nuclear spins.  Qubit initialization and readout is performed by a
combination of transfer of spin between nuclear and electron spins (for
example by using the SWAP operation) and measurement operations on the
electron spins.

\begin{figure}
\vspace{-1.5cm}
\begin{center}
\includegraphics[width=8.5cm,angle=0]{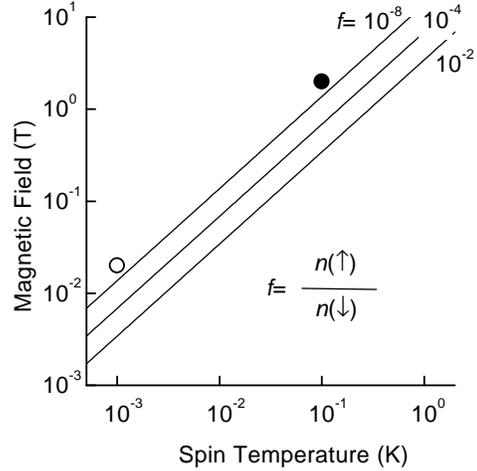}
\end{center}
\vspace{-3cm}
\caption{Contours of 
equilibrium electron spin polarization plotted as a function
of $B$ and temperature $T$.  Extremely
high spin polarization requires a combination of very low
$T$ and large $B$.  While $T$=100 mK and $B$=1 T (filled circle) 
may readily be obtained in the laboratory,
operation at lower magnetic fields and lower spin
temperatures may be possible (open circle) if on-chip spin refrigeration
devices can be developed.}
\end{figure}

\section{Environment Necessary for Quantum Computation}

As was mentioned above, the long relaxation times of Si:P necessary
for quantum computation only occur at $T \leq$ 4 K.  At higher
temperatures phonon scattering rapidly decreases the relaxation times
\cite{Feher59b}\cite{Honig60}.
A more stringent requirement placed on the environment necessary for
quantum computation is that the spins must be highly polarized.  The
equilibrium
polarization of electrons as a function of $T$ and $B$ is plotted in
Fig. 2.  Strong polarization of the electrons requires both high
magnetic fields ($B \simeq$ 1 T) and very low temperatures
($T \simeq$ 0.1 K).  These conditions were assumed in the original
quantum computer proposal and are readily obtainable in low
temperature experimental laboratories.  The extremely long relaxation
times of the spins imply that $nonequilibrium$ polarizations of the
spins are possible in less extreme environments, achieved for
example by injection of electrons from ferromagnetic contacts or
by optical pumping. I will discuss below a third alternative
for polarizing electron spins by using devices which perform
$spin~refrigeration$.  These ideas, if implemented, would enable
a spin quantum computer to operate in the more favorable environment
of higher temperatures and lower magnetic fields.

\begin{figure}
\vspace{-1.5cm}
\begin{center}
\includegraphics[width=8.5cm,angle=0]{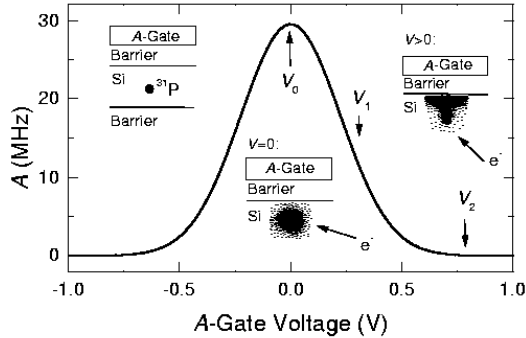}
\end{center}
\vspace{-3.5cm}
\caption{An $A$-gate is located above P donors in Si and
controls the hyperfine interaction between the donor nucleus
and the surrounding electron by distorting the electron
wave function.  The plot is a qualitative depiction of $A$
as a function of gate voltage.  $A$ is least sensitive to
gate voltage fluctuations when $dA/dV=0$ (at $V_0$ and
$V_2$) and most sensitive to voltage fluctuations near
$V=V_1$.}
\end{figure}

\section{Gate-Controlled Quantum Logic}

The Si:P two qubit quantum computer cannot be integrated into
a large-scale quantum computer unless quantum operations can
be applied selectively to particular spins (bulk spin resonance
operations perform identical operations on all spins in the
system).  Also, connectivity must be established between spins
in order to do arbitrary multi-qubit quantum logic.  Both
selectivity and connectivity can be achieved by applying
electric fields to metal gates adjacent to the spins.
An approach to selective single spin quantum logic is shown
in Fig. 3. A metal ``$A$-gate" is located directly above a
donor site, separated by a barrier from the Si semiconductor
material (possible materials that could be used for the
barrier will be discussed below).  Application of a voltage
bias to the gate creates an electric field which distorts the
electron wave function surrounding the donor \cite{Valiev99}.  This distortion
changes the electron density at the donor nuclear site and also
the hyperfine interaction energy $A$.  Because the energy
spacings of the spin levels are sensitive to $A$, an $A$-gate
can selectively bring a single Si:P system into (or out of)
resonance with a globally applied external $B_{rf}$.  Single
spin quantum logic can thus be performed only at selected
sites.

\begin{figure}
\vspace{-1cm}
\begin{center}
\includegraphics[width=8.5cm,angle=0]{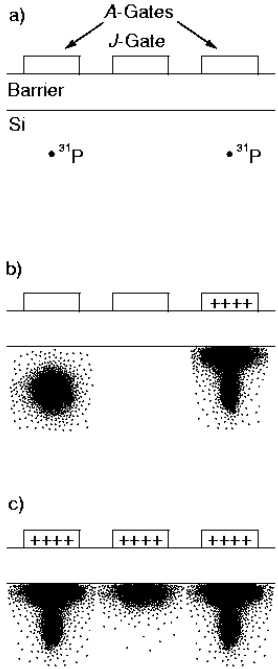}
\end{center}
\vspace{0cm}
\caption{(a) Configuration of gates and two donors for doing one
and two qubit logical operations: in addition to $A$-gates 
above the donors, $J$-gates lie between donor sites.  (b) One
qubit logical operations are performed by applying an $A$-gate
bias which brings a selected spin into resonance with an
external rf magnetic field.  (c) Two qubit operations are
performed by lowering the potential barrier between donor
sites with the $J$-gate and turning on exchange coupling
between the donors.  Electron mediated nuclear spin exchange
will then occur between the donor nuclei.}
\end{figure}

The simplest way to couple additional spins into a quantum computer
architecture is to fabricate an array of Si:P sites in close proximity
(Fig. 4).  As the site separation becomes comparable to the Bohr radius,
the electron wave functions begin to overlap.  Tunneling of electrons
between sites then becomes possible, leading to an exchange
interaction between the electron spins and also to an indirect (or electron
mediated) exchange interaction between the nuclear spins.
For the case when the two electron spins are in the $|\dd \rangle$
state
and are each coupled to the donor nuclear spins by the same
hyperfine interaction energy $A$, the nuclear spin exchange frequency
is approximately: 

\begin{equation}
h \nu_J= 2 A^2 (\frac{1}{ \mub B -2J}-\frac{1}{ \mub B} )  ,    \label{5}
\end{equation}
where $J$ is the electron spin exchange energy, and $2J<\mub B$ is
assumed.  The magnitude of $J$ between electron spins
on donors as a function of their separation $r$
can be approximated from equations derived for the case of well-
separated H atoms \cite{Herring64}:

\begin{equation}
J (r) \sim  E_b (\frac{r}{a_B})^\frac{5}{2}
\exp (\frac{-2r}{a_B}) .   \label{4}
\end{equation}
This function, with values appropriate for Si, is plotted in Fig. 5.
Substantial nuclear spin exchange between donors requires that
$J$ and $\mub B$ must be comparable, and consequently the
donors and gates must be spaced of order 100 \AA~ apart, a scale
which is near the limit of current nanofabrication technology.

\begin{figure}
\vspace{-2cm}
\begin{center}
\includegraphics[width=8.5cm,angle=0]{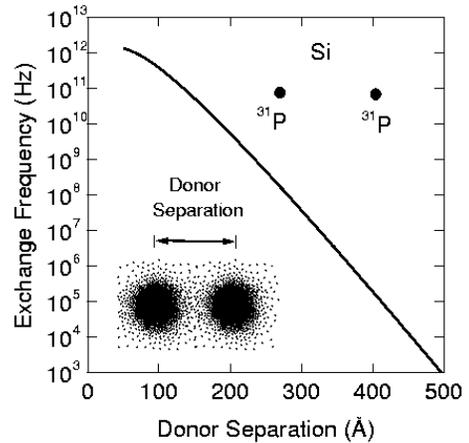}
\end{center}
\vspace{-2cm}
\caption{The strength of the electron exchange coupling between donor
sites, plotted as a function of their separation.  The exchange
coupling must be comparable to the electron Zeeman energy for
nuclear spin exchange to occur between sites.  This requirement
means the donor spacing must be of order 100 \AA.}
\end{figure}

Because exchange interactions depend on the overlap of the electron wave
functions, they can be effectively controlled by a voltage bias applied
to a ``$J$-gate" lying $between$ donor sites (Fig. 4). Exchange interactions
lead naturally to the SWAP operation of quantum logic, in which the spin
quantum numbers of two qubits are interchanged.  The SWAP operation in
combination with single spin operations can be used as the primitive
operations of a universal quantum computer \cite{Loss98}.  In the configuration shown
in Fig. 4, a negative voltage bias applied to the $J$-gate decouples
the adjacent spins. A positive voltage applied to the gate turns on
the exchange interaction between the spins.  A SWAP operation is performed
if the voltage pulse has the appropriate length.  Calculating the value
of the exchange interaction as a function of separation between donors
and of gate voltage in realistic device configurations is difficult and is the 
subject of current research \cite{Burkard99} \cite{Hu99}.
An added complication in Si is that the degenerate band structure leads
to oscillatory behavior of the exchange interaction due to interference
between electrons in different conduction band minima \cite{Andres81}.

\section{Speed of Computer Operation}

The speed of the computer architecture described above is limited by
the rate at which single spin operations can be performed.  This rate
is approximately the product of $B_{rf}$ and $\mu_{eff.}$, plotted in Fig. 1c.  
$B_{rf}$ cannot be too large or else
it will excite transitions between states not resonant with the field.
Additionally, large values of $B_{rf}$ will introduce eddy current heating
of conductors (gate leads for example) in the neighborhood of the computer.
A reasonable value of $B_{rf}$ is $10^{-3}$ T.  This value, and a $\mu_{eff.}$
appropriate for nuclear spin flips at $B$=1 T, yields a single spin operation rate of
$r \approx$10-100 kHz.  

The speed of two spin operations will depend on the strength of the exchange
coupling between sites, but can be comparable to or greater than the rate for
single spin operations.  More relevant is that spins can only interact with
their neighbors, so that many SWAP operations must be performed to do a
two spin operation on two spins that are far apart.

It is certainly possible to find alternative systems in which gate operations
could be performed much more rapidly than the system discussed above.
The stringent requirements on the accuracy of logical operations in a quantum
computer are likely to be more easily fulfilled in systems with slow dynamics, so the
slowness of logical operations of the nuclear spin quantum computer is not
necessarily a disadvantage.  Clearly, it is the ability to perform quantum
algorithms, not the clock speed, which will distinguish any quantum
computer from its classical counterparts.

\section{Noise Introduced by Gate Voltage Fluctuations and Gate Calibration}

More relevant than the speed of a quantum computer is the $ratio$ of the
time required for logical operation to the decoherence time of the qubits.
If this ratio is less than $\sim 10^{-4}$, then perfect error correction
becomes possible \cite{Gottesman99} \cite{Knill98}.
The long relaxation times noted above for Si:P were measured in bulk samples
with very low doping density.  In the computer architecture, however, the
spins are located near a surface and beneath metallic gates, which will
inevitably introduce additional decoherence mechanisms for the spins.  While
the degree of degradation of the spin relaxation rates will need to be determined
by experiments, it is worthwhile to estimate the decoherence introduced by
thermal fluctuations of the gate voltages on the spins.  The simplest case
to treat is the effect of voltage fluctuations on $A$-gates on the spin beneath
them.  The spin is essentially a voltage controlled oscillator, and voltage
fluctuations lead to phase errors (Fig. 6).  For a white noise spectrum of
voltage fluctuations with spectral density $S_V$ the dephasing rate is:

\begin{equation}
\tfi^{-1} = \pi^2 \alpha^2 (V) S_V ,    \label{7}
\end{equation} 
where $\alpha \equiv d \nu / d V$ is the tuning parameter of
the VCO.  From Fig. 3, it is clear that $\alpha$ can be arbitrarily
small if the $A$-gate is biased appropriately.  To obtain a crude
estimate of $\tfi^{-1} $, I assume $\alpha$ = 100 MHz Volt$^{-1}$.  The Johnson
noise on a 50 $\Omega$ transmission line at room temperature is
$S_V \cong 10^{-18}$ V$^2$Hz$^{-1}$.  These values give $\tfi^{-1} $=0.1 sec$^{-1}$, 
allowing many thousands of logical operations to be performed in the
decoherence time of the computer, and indicating that Si based nuclear spin
quantum computers are potentially in the realm where continuous error correction is
possible.

\begin{figure}
\vspace{-1cm}
\begin{center}
\includegraphics[width=8.5cm,angle=0]{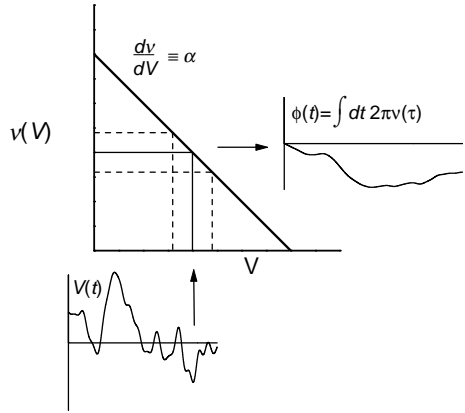}
\end{center}
\vspace{-3cm}
\caption{Since gate voltages are used to control the dynamics
of qubits in the proposed architecture, fluctuations in gate
bias must inevitably be a source of decoherence.  An $A$-gate
over a donor can be modeled as a classical voltage-controlled oscillator.
The larger the VCO tuning parameter $\alpha$ the more rapidly
phase error in the spin state builds up with time.}
\end{figure}

A more serious issue arises from low frequency fluctuations ($1/f$
noise) of voltages on the gates and of electric fields at the
donors arising from charge motion within the semiconductor host.
As $f \rightarrow 0$ these fluctuations become variable offsets
of the gates that must be calibrated for the computer to operate
properly.  In principle each of the gates in the computer can
be individually calibrated by performing simple measurements on
the operation of each gate.  Clearly, however, this process will
become increasingly cumbersome as the size of the computer is scaled up,
particularly if regular recalibration is necessary.
A computer architecture which does not require calibration of
individual gates, and is consequently resilient in the presence
of low frequency fluctuations, is discussed below.

\section{Techniques for Single Spin Measurement}

The procedures outlined above perform all of the logical operations
necessary for quantum computation.  The spins must also be prepared
in a specified initial state and read out at the end of the calculation.
In a computer architecture capable of performing error correction,
state preparation and measurement are necessary
throughout the process of calculation.  For this to be possible the
measurement must be rapid, since if the measurement process
during error correction were slow, the qubits would decohere during
the period in which the measurement was occurring.  While the
measurement of the magnetic fields induced by electron and even
nuclear spins may be possible using advanced magnetic resonance
force microscopy \cite{Sidles95} \cite{Wago97},
these techniques are highly unlikely to perform
measurements on a time scale comparable to logical operations.
While single spin measurements are likely to be both difficult
and slow, single charge measurements on a microsecond time scale
are now routinely performed using single electron transistor
(SET) electrometers \cite{Schoelkopf98},
and SET's may be ideally suited to perform
the quantum measurements necessary for quantum computation
\cite{Shnirman98}.  

Probably the simplest system in which spin to charge conversion
is possible is a system of two electrons in a common potential
well.  The Pauli Principle forbids both electrons from being in the
same orbital state unless they have opposite spins and are in 
a singlet state.  Consequently, in the absence of a magnetic
field the lowest energy state of a two electron system is a
spin singlet, and triplet states are higher in energy
\cite{Ashcroft}.  An
electrometer capable of detecting the number of electrons
occupying a bound state can thus determine whether two electrons
are in a singlet or triplet state. 

Measurement of a single electron spin requires that one of the
two electrons be in a known spin state, for example,
$|\downarrow ? \rangle$.  (Techniques for preparing electrons
in known spin states will be discussed below.)  If the state
of the system is $|\dd \rangle$ then a measurement of the two
electrons will yield a triplet result.  If the two electrons
are in the $|\ud \rangle$ state, then they
may be either in a singlet ($\ss$) or a triplet ($\st$) state.  
Scattering between $\ss$ and $\st$ 
is generally much more rapid then scattering between
$|\dd \rangle$ and $|\du \rangle$, since scattering between the latter states
can only proceed via a spin flip.
Consequently, a measurement which determines whether the electrons
are in a singlet or triplet state can be used to infer the
spin state of the second electron if the measurement time
exceeds the time required for scattering between
the  $\ss$ and $\st$ states but is less than the time
required for scattering between $| \dd \rangle$ and $|\du \rangle$.

\begin{figure}
\vspace{-2cm}
\begin{center}
\includegraphics[width=8.5cm,angle=0]{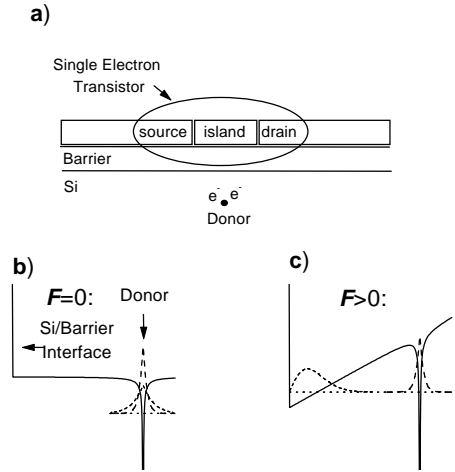}
\end{center}
\vspace{-3cm}
\caption{(a) Possible configuration in which a
single electron transistor (SET) is used as a sensitive
electrometer to probe the spin state of a two-electron system.
The conduction across a SET depends sensitively on the potential
of the island electrode.  (b) In the absence of an applied electric
field $F$ the wavefunctions of both electrons (dotted lines) 
are localized by the attractive electrostatic potential (solid
line) in the vicinity of the donor.  (c) An applied electric field
can draw one of the two
electrons to a state at the Si-barrier interface.  The value
of $F$ where this electron moves from the interface to the donor
is different for electrons in singlet and triplet spin states.
This charge motion affects the SET conductance, enabling
the spin state of the electrons to be determined.}
\end{figure}

The potential well binding the two electrons can be provided by
a donor state in Si.  A singly charged donor site, such as Si:P
is capable of binding two electrons to form a $D^-$ state only
if the electrons are in a singlet state.  The binding energy of
this state is small, however, and probing it will be difficult
\cite{Larsen81} \cite{Larsen92}.
Doubly charged donors, such as Column VI impurities in Si, have
strongly bound two electron states.  In Si:Te, the singlet ground
state is almost 150 meV below the lowest lying triplet state
\cite{Grimmeiss86} \cite{Grossmann87}.  Column VI donors also
have stable isotopes with zero nuclear spin, allowing two
electron systems to be probed in the absence of an additional
hyperfine interaction with the donor nucleus.

An experimental configuration in which a two electron system is
probed by a SET is shown in Fig. 7.  The SET island lies directly
above a donor embedded in Si, which is separated from the SET by
a barrier layer.  A potential bias between the island and the
substrate is capable of ionizing the donor and drawing one of
the two electrons to a state at the Si/barrier interface.
Charge motion between the interface state and the donor state
will change the potential of the SET island and hence the
conductance of the SET.  Because the binding energy of the
donor is different for singlet and triplet spin states,
detection of charge motion onto the impurity will
allow the spin state of the electrons to be
determined.

For an rf-SET \cite{Schoelkopf98},
well coupled to a Si:Te donor, a measurement time
of order microseconds has been estimated \cite{Kane00}, much
smaller then the expected spin flip scattering time
($|\dd \rangle \leftrightarrow |\du \rangle$) of the system.
It may also be possible to measure the spin state
using a more conventional FET as an electrometer \cite{Vrijen99}.
Clearly experiments will be necessary to determine
the actual speed of measurement and the scattering
times in real systems.

Because electron and nuclear spins
are coupled by the hyperfine interaction, polarization
transfer between electrons and nuclei are possible.
Consequently, a nuclear spin can be measured by
allowing it to interact with a two electron system
in a known spin state, and subsequently measuring
the electrons using the procedure discussed above.  

\section{Possible Improvements to the Architecture}

The quantum computer architecture outlined above shows
that quantum computation and single spin measurement
in Si nanostructures are
theoretically possible using devices that are not
extraordinarily different from those being currently
fabricated.  Nonetheless there are several deficiencies
of the architecture that may prevent it from being
a practical one, and several improvements on the
original design have been suggested. I will focus
below on some obvious shortcomings: (1) the computer
must operate at extremely low temperatures and in
very high DC magnetic fields, (2) quantum logic
is only possible between nearest neighbor spins,
and these spins must be extremely close together
($\sim 100$ \AA) for the computer to operate, and
(3) logical operations must be performed with a
precision far exceeding that which is typically
obtainable in solid state devices in order to
make continuous computation possible.  I will
discuss below possible modifications to the
original architecture which may alleviate the
difficulties associated with each of these
problems.

\subsection{Spin Refrigeration}

While the computer must operate at extremely low
temperatures and in high magnetic fields to
fully polarize the electrons,
a more moderate environment ($T$=1-4 K) would vastly simplify the
construction of a computer with many components,
when power dissipation will inevitably become a factor.
Operation at smaller magnetic fields would also
be desirable, firstly since many conventional
semiconductor components, which may need to be
placed near the computer (or even on the same Si
chip), are rendered inoperative
in large magnetic fields.  Secondly, matrix elements
connecting hyperfine-coupled nuclear spin states
increase with reduced magnetic field (Fig. 1c) and
operation of the computer can potentially proceed
more rapidly at lower magnetic fields.  Finally,
altogether different quantum computer architectures
may be possible as $B \rightarrow 0$.

While $T$=100 mK is necessary to fully polarize the
electron spins in large laboratory magnetic fields, the spin-
lattice relaxation time is extremely long even at
$T \approx$4 K, where $t_1$ of electrons on donors still exceeds
1 sec \cite{Feher59b} \cite{Honig60}.
Consequently it is relatively easy to create
a situation where the spin temperature is much different
from the lattice temperature.  Construction of a
`spin refrigerator' would create highly polarized spins
at more modest lattice temperatures and magnetic fields.

Possible methods to introduce nonequilibrium polarizations
of the electrons include injection of spins from ferromagnetic
contacts and optical pumping.  Neither of these techniques
alone, however, is likely to achieve the extremely high
polarizations necessary for large-scale quantum computation.
What would be desirable is an on chip, closed cycle, spin
refrigeration device which could increase electron spin
polarization.

\begin{figure}
\vspace{-1cm}
\begin{center}
\includegraphics[width=8.5cm,angle=0]{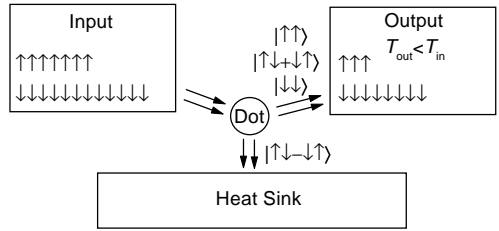}
\end{center}
\vspace{-3cm}
\caption{Devices which can sort singlet and triplet states of
electron pairs can be used to refrigerate spins.  In the
figure, randomly selected pairs of electrons from a partially
polarized input reservoir pass through a dot.  Singlet pairs
are expelled into a heat sink, while triplet pairs pass through
to an output reservoir.  Because the polarization of the singlet
pair is exactly zero, the polarization of the output
reservoir must be greater than in the input reservoir and
consequently $T_{out} < T_{in}$.}
\end{figure}

A possible realization of such a device using the ideas of
single spin measurement discussed above is depicted in Fig. 8.
Electrons in a partially polarized `input' reservoir proceed
in pairs through a `dot' and into an `output' reservoir.
On the dot a measurement is performed which distinguishes
triplets from singlets.  Triplet states are passed through
the dot into the output, while singlet pairs are expelled
into a heat sink.  Because singlet pairs have exactly
zero spin polarization, the output reservoir must have higher
average polarization than the input reservoir, and hence
lower temperature.  Such `singlet rejection refrigeration
devices' could be cascaded to produce the high polarizations
necessary for quantum computation starting from a modestly
polarized input, created either thermally or by injection
from a ferromagnet.  A circulating bath of cooled polarized
electrons could polarize the nuclei of a quantum computer
via the hyperfine interaction to initialize qubits
and provide the ancilla
spins necessary for error correction and measurement.

Very recently, an approach to  quantum computing
has been developed based solely on controlling
exchange interactions between spins to perform logical operations
on qubits encoded on several spins \cite{Bacon99}.  Using this approach
quantum operations would not require magnetic fields, and
spin refrigeration would enable the computer to operate
at $B \rightarrow 0$.

\subsection{Use of Electron Spins as qubits for transport of
quantum information}

In the quantum computer architecture presented above
nuclear spin qubits are coupled to
each other via indirect exchange processes mediated by the
electrons, and the spatial separation between spin qubits
cannot greatly exceed the Bohr radius of the donor impurities.
Recent experiments \cite{Kikkawa99} clearly show, however, that electron
spins can effectively transmit quantum information over
much larger distances ($\simeq 100 \mu m$).  While it is important to note
that a free electron at a Si interface will have altogether
different (and generally shorter) relaxation times than those
for an electron bound to a donor, transmitting quantum
information using a free electron \cite{Loss99} has the obvious
advantage that the donor nuclei being coupled can
be separated over much larger distances than the
Bohr radius.  Quantum information transport by free
electrons could also enable quantum computers to have separated,
specialized devices for logic, memory, and measurement.
Thus, for example, Te donors below SET's could
be optimized for measurement operations, while P
donors below $A$-gates could be optimized for quantum
logic.

\begin{figure}
\vspace{-2cm}
\begin{center}
\includegraphics[width=8.5cm,angle=0]{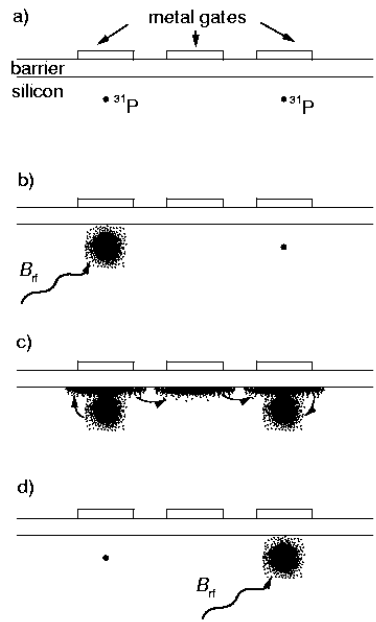}
\end{center}
\vspace{0cm}
\caption{(a) Possible architecture for quantum logic
in which quantum information is transmitted by free electrons.
(b) An rf pulse performs an operation which swaps the spin state
of the nucleus and the donor electron. (c) The electron is then
ionized from the donor and shuttled to a different donor site
by applying biases to gates in a manner similar to that used
in a charge coupled device. (d) Once it is
bound to the second donor, a second rf pulse exchanges spin
between the electron and the donor nucleus, completing transfer
of quantum information between spins.  In this architecture
donor spacing can be much larger than in approaches which
use exchange coupling between donors for quantum information
transfer.}
\end{figure}

A gated structure in which quantum logic
between nuclear spin qubits is mediated by
a free electron spin is shown in Fig. 9.  First, quantum
information is swapped between a donor nuclear spin and 
an electron bound to the donor by an appropriate rf
pulse.  The electron is then ionized from the
donor by an electric field applied to the gate above
the donor and is transmitted to a second
donor site through one or more intermediary gates which
move the electron along the interface.  The electron is
then allowed to combine with the second donor 
and a second SWAP operation
is applied to exchange information between the electron and
nuclear spin.  While the architecture shown in Fig. 9 is
similar to that depicted in Fig. 4, the gate dimensions
need not be comparable to the donor Bohr radius.

\begin{figure}
\vspace{-2cm}
\begin{center}
\includegraphics[width=8.5cm,angle=0]{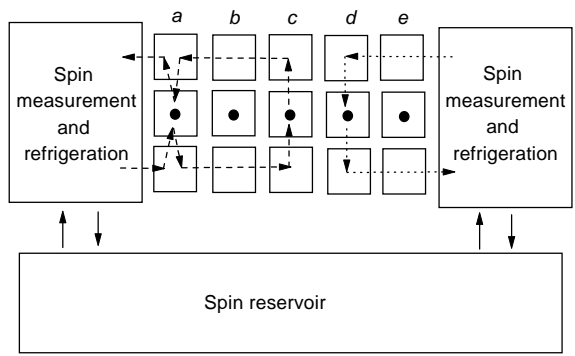}
\end{center}
\vspace{-3cm}
\caption{Drawing of a five qubit quantum computer using gates
(squares)
to transfer unbound electrons between nuclear spin qubits
(black circles).
Measurement/refrigeration devices are adjacent to the qubits.
Dashed path is for an electron which performs a two qubit
logical operation on spins $a$ and $c$.  An electron following
the dotted path performs a measurement operation on spin $d$.
Note that all donor sites and gates do not need to be
functioning to do quantum operations on the computer.}
\end{figure}

Quantum information in this scenario must be transmitted
through a single electron spin.  If the mediating electron
couples to other electrons through the exchange interaction,
then quantum information will be lost.  Thus the gate
electrodes moving the electron from site to site must
store a single electron at a time, and the architecture
strongly resembles a single electron charge coupled device
(CCD) \cite{Sze}.  Also, the interface along which the
electron is transmitted must be almost completely
free of trapped charges.
A possible quantum computer architecture using these
types of devices is shown in Fig. 10.  Gate electrodes
shuttle single electrons between donors to perform
quantum logic and between donors and electron spin
measurement devices for measurement of the qubits.
The measurement devices also provide a source of
polarized spins for qubit initialization and
error correction.  Such an architecture can be made
defect tolerant \cite{Heath98}, since all gates and donor sites do
not need to be functional in order to do quantum
operations and measurement. 

Using electron spins to transmit quantum information
over relatively large distance will have obvious
advantages in terms of implementing quantum algorithms.
While quantum computation \cite{Benjamin00}
and error correction \cite{Gottesman99}
are possible using only local gates, 
efficiency will undoubtedly
improve if quantum information can be rapidly transmitted between
remote qubits.

\subsection{Modified RF pulses for uniform gate operations}

Decoherence can be produced in a quantum system not only
by interactions of the qubits with uncontrolled degrees
of freedom but also by errors in the
logical operations on the qubits.  Consequently, the accuracy of
gate operations in a quantum computer will need to be
$\sim 10^{-4}$ to be in the regime in which error correction
will enable continuous quantum computation.  This level of
precision will be an extremely difficult requirement to meet
in any solid state material, where fluctuations of device
properties are inevitable.  One possible solution is to
perform precise calibration of each individual gate in the
computer, an approach which will become increasingly difficult
in large computers and will be even more arduous if the
calibrations themselves fluctuate.  A more attractive
possibility pointed out by Benjamin \cite{Benjamin00} is to design gate
operations which are highly precise despite device fluctuations.

\begin{figure}
\vspace{-1cm}
\begin{center}
\includegraphics[width=8.5cm,angle=0]{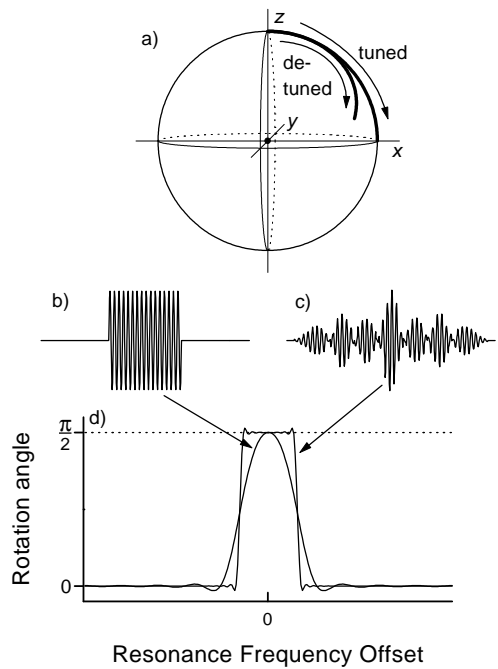}
\end{center}
\vspace{-1cm}
\caption{(a) Illustration of the Bloch sphere of a single qubit.
An rf pulse (b) that performs a $( { \pi \over 2 } )_y$ operation on
a qubit resonant with the pulse frequency will perform a
different operation on a nonresonant qubit.  Modification of
the rf pulse shape (c) can compensate for these errors in the
neighborhood of the resonance to very high accuracy, however (d).}
\end{figure}

An example of a fluctuation-tolerant approach to single spin
rotations is shown in Fig. 11.  A rotation induced
by an rf pulse will deviate from the desired value if the
resonance frequency of the qubit deviates from the frequency
of the pulse.  If the rf pulse is shaped appropriately,
however, the errors associated with small resonance frequency
offsets can be drastically reduced. Shaped and composite
pulse waveforms for single spin rotations have
been developed to improve accuracy of NMR spectroscopy 
\cite{Geen93} \cite{Freeman} and
of bulk spin resonance quantum logic \cite{Cummins99}.
Extension
of these ideas to two qubit operations - while presumably
possible in principle - has not yet been addressed.  In the
computer architecture shown in Fig. 9, different types of
rf pulses are used to perform both one and two qubit operations,
and shaped pulses will allow accurate logical operations to be
performed despite appreciable device fluctuations.  Because
the effect of a shaped pulse is independent of the exact
resonance frequency of the qubit, low frequency fluctuations
of the resonance frequency would not degrade
the accuracy of the logical operations of the computer,
and the computer could consequently be highly immune
to the inevitable 1/$f$ noise present in solid state
materials.

Another problem with the original architecture is that each
gate is a channel coupling to the system that can be a source of
decoherence \cite{Benjamin00}.
While this decoherence is an inevitable byproduct
of using gates to control logical operations, the magnitude of
the decoherence varies with applied gate bias.  For example in
Fig. 3 gate-induced decoherence is largest when $\alpha$ is
largest, near $V=V_1$, while it is much smaller when 
$\alpha \rightarrow 0$ at $V=V_0$ and $V=V_2$.  Shaped pulses
could enable ``digitally gated" quantum computer architectures
in which the gates are biased only at two low decoherence
values, one resonant and one nonresonant with the applied
$B_{rf}$. Gates associated with idling qubits are biased at
their nonresonant value.  A logical operation on a spin is
performed by rapidly bringing the gate above the spin
to its resonant voltage
and applying an rf pulse.  At the completion of the pulse,
the gate is rapidly brought back to its non-resonant setting.
In this manner the gates still allow selective operations to
be performed on the qubits without introducing decoherence,
except during the brief periods when the gates are switched.

Even if shaped pulses are used for quantum logic,
it is likely that some degree of calibration will be necessary
in any solid state quantum computer.  While this may appear
to be a fundamental obstacle to large-scale quantum computation,
it is important to remember that the classical computational
resources associated with the operation of any quantum
computer can be large, and consequently a large amount of
conventional computer memory and processing power can be
devoted to the proper calibration of each gate.

\section{Implementing the Architecture in \si/\sige~ Heterostructures}

In all the device structures discussed above, a barrier material
must be present separating the Si containing the donors from the
conducting gates. The Si/barrier interface must be almost entirely
free of charge and spin defects if the devices are to perform
quantum gate operations, especially if free electrons on the interface
are to be used to transfer quantum information between remote donors.
Silicon oxide and nitride layers - used
in conventional MOS structures - are amorphous materials, and
their interfaces with Si will inevitably contain charge
centers associated with dangling bonds, rendering their use
in a quantum computer highly problematic \cite{Lucovsky99}.
Various epitaxial
barrier materials on Si have been developed, however, that may ultimately
have the extraordinarily low defect densities that will
be necessary in a quantum computer.

Several oxide materials, including CeO$_2$
\cite{Jones98} and SrTiO$_3$ \cite{McKee98}, are being
explored which may grow epitaxially on pure Si. 
Most promising for application to quantum computation, however,
are the heterostructures grown with Group IV elements:
Si/\sige~ \cite{Vrijen99} and Si/\sic. I will focus on
Si/\sige~ heterostructures because they are the most
technologically developed.  Si/\sic~ heterostructures
have recently been used to perform the first direct 
electron spin resonance measurements \cite{Nestle97} \cite{Kummerer99} 
on two dimensional electron systems (2DES), and they
may thus also be relevant for implementing quantum logical devices.

\begin{figure}
\vspace{0cm}
\begin{center}
\includegraphics[width=8.5cm,angle=0]{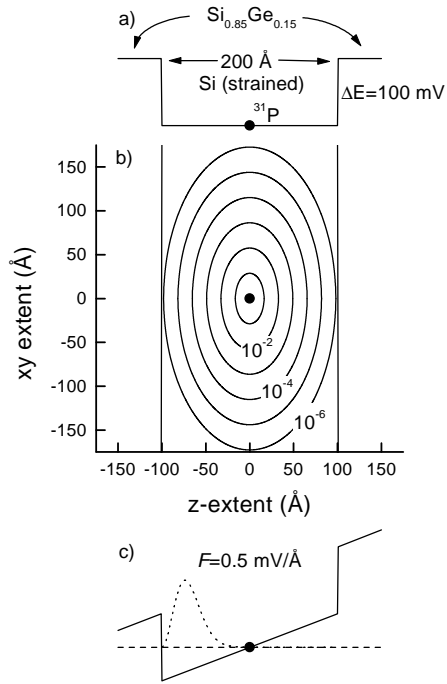}
\end{center}
\vspace{-1cm}
\caption{(a) Possible realization of the quantum computer architecture
where the electron and the donor reside in a strained Si quantum
well surrounded by unstrained \sige. (b) Because of the strain, the
electron mass for motion parallel to the layers is smaller than the mass
for motion perpendicular to the layers, and the contours of
equal probability of the electron wavefunction are ellipsoidal.
(c) An
applied electric field $F$ can draw the electron to an
interface state confined in the well.}
\end{figure}

Because of the lattice mismatch between Si and Ge, \sige~
heterostructures must inevitably contain layers that are
strained \cite{Jain}.  
A high quality (mobility 500,000 \mob \cite{Ismail95})
2DES has been formed in these
materials by confining the electrons in a strained Si
layer between unstrained layers of \sige~ on a
(100) oriented substrate (Fig. 12a).
Because the Si is strained along a (100) axis, the valley degeneracy of
the conduction band is broken, and the well is occupied
by electrons only in the two valleys with minima on the axis
perpendicular to the layers.
The electron effective mass in
these valleys is anisotropic \cite{Ando82}, and the contours
of equal probability density of electron
wavefunctions on donors in the Si well are ellipsoidal (Fig. 12b).

Since the Si layer is under strain, its thickness cannot
exceed a critical value without the nucleation of dislocations \cite{Jain}.
For Si/Si$_{0.85}$Ge$_{0.15}$ this thickness is approximately 200 \AA,
sufficiently thick so that the electron wavefunction of
a donor in the center of the well is almost entirely
in pure Si. 

\begin{figure}
\vspace{0cm}
\begin{center}
\includegraphics[width=8.5cm,angle=0]{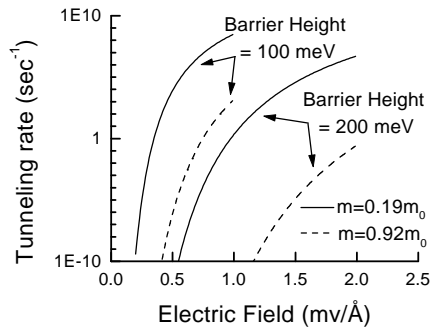}
\end{center}
\vspace{-3cm}
\caption{Fowler-Nordheim tunneling rates for electrons out of
a strained Si quantum well through a \sige~ barrier.
Si/\sige~ structures for quantum
logic will need to be carefully designed so that electron
tunneling out of the well is not a significant source of
decoherence.}
\end{figure}

Conduction band offsets in Si/\sige~ are not large,
and consequently electron tunneling through the
barrier is a potential source of decoherence.
Estimates of the Fowler-Nordheim tunneling rate
\cite{Nagano94} depend strongly on the electron
mass, however (Fig. 13).  While tunneling out of the well
is in the direction of the heavier Si electron mass,
coupling between valleys may be large enough that
tunneling rates will be intermediate between 
those calculated for the
two electron mass values.  Experiments will be
necessary to determine electron tunneling rates
before Si/\sige~ devices for quantum computing
can be optimized.  For $x$=0.85 the depth of the well is about
100 meV, probably sufficient to allow the application of
large enough electric fields
to ionize the donor (Fig. 12c) without excessive
tunneling leakage.  

\begin{figure}
\vspace{0cm}
\begin{center}
\includegraphics[width=8.5cm,angle=0]{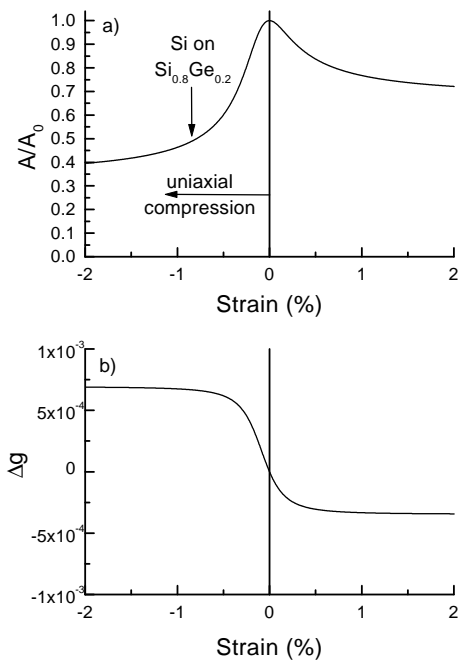}
\end{center}
\vspace{-1cm}
\caption{The effect of strain on the hyperfine interaction
frequency (a) and the $g$-factor anisotropy (b) in Si,
using calculations in Ref. \protect\cite{Wilson61}.
In the strained layer stucture shown in Fig. 12, the
hyperfine interaction will be reduced about 50\% from
the value observed in unstrained Si.}
\end{figure}

Straining
the Si layer also reduces the hyperfine interaction
energy $A$ and the $g$-factor anisotropy of the coupled
electron-nuclear spin states (Fig. 14) \cite{Wilson61}.
In the strained well $A$ will be approximately 50\%
of its value in unstrained Si.  Also
relevant is that fluctuations
in static strain (created during device fabrication for
example) will affect the values of the resonance frequencies.
Fluctuations of the $g$ factor caused by strain are actually
reduced in highly strained layers, however.

\section{Dopant Introduction by Low Energy Ion Implantation}

Finally, I consider possible approaches to introducing single
donor ions in controllable locations into the Si.  The low density
of unwanted impurities introduced into the Si that is required
for the computer suggests that both crystal growth and
dopant incorporation should be performed in 
ultra high vacuum in the same apparatus.
Possible approaches to dopant placement include nanoassembly
using a scanned probe and ion implantation \cite{Vrijen99}.  At the
temperatures necessary for optimal growth of high quality
material \cite{Ismail95}, surface segregation of donors 
deposited onto a
Si surface is substantial \cite{Hobart96},
leading to large ($\geq 100$\AA) vertical
(and presumably lateral) displacements of deposited donors
from their original location.  The diffusion of donors
once they are incorporated into the bulk is, however,
much lower.  Low energy ($\simeq$150 eV) 
ion implantation has consequently
been developed as a tool for introducing donors into
semiconductors with extremely sharp ($\leq 30$\AA )
vertical profiles \cite{Fons88}.
A tool which could place donors into Si with the
precision necessary for development of the proposed
quantum computer would be a low energy focused
ion beam with single ion implantation capability \cite{Shinada98}
incorporated into a Si MBE growth chamber.  The development
of such a tool would likely be a significant technological
challenge, and many important experiments can first be done to
assess the viability of the proposed architecture with
simpler techniques and equipment.

\section{Prospects and Conclusions}

The obstacles
to building a silicon-based quantum computer are
enormous, and much research needs to be done before even the
viability of the ideas presented above can be
determined.  Priorities in this research must be:
(1) the fabrication of single spin measurement
devices, (2) measurement of spin decoherence mechanisms of
electrons and nuclei at donors and of free electrons
in real Si heterostructure materials, (3) the development
of single charge CCD-like devices capable of transporting
quantum information on an electron spin, (4) the design of
quantum logical operations which are tolerant to the
fluctuations inevitable in solid state materials, and
finally (5) the demonstration of a technology for placing
single donors into a semiconductor substrate at precisely
specified locations.

Single spin measurement devices (and also presumably spin
refrigerators) will be of interest in their own right,
irrespective of their impact on quantum computation.
It is likely that they can be fabricated with currently
available technology \cite{Kane00},
and they can subsequently be applied to the
detailed measurement of decoherence in real nanostructure
devices.  The development of single charge CCD's  will
be an important milestone showing that the extreme
material purity necessary for the proposed quantum
computer architecture is
achievable.

Large-scale quantum computers will need to have fluctuation
and defect tolerant designs, both because of the precision that
is required for doing quantum logic and because of the
inevitable variations in solid state materials from device
to device.  The technology for donor placement which will
need to be developed may also have applications outside of quantum
computation, since fluctuations attributable to the random
locations of donors are already an issue for ultra-small
conventional semiconductor devices.

Because technological advances need to be made on many
fronts, it is currently not possible to ascertain whether the
architecture for Si-based quantum computers presented
above is viable.  The outcome of research in the next few years
focused on the issues mentioned above, however,
will determine whether silicon - the material used in today's
computers -is also capable of being the host
material for a revolutionary new type of machine.

\section{Acknowledgements}

The author has benefited from discussions with S. Barrett,
X. Hu, 
G. Lucovsky, S. Lyon, J. Melngailis, R. Schoelkopf,
P. Thompson, and E. Yablonovich.

\end{document}